\documentclass[12pt]{article}

\usepackage{geometry}             
\usepackage{graphicx}
\usepackage{amssymb}
\usepackage{amsmath}
\usepackage{epstopdf}

\usepackage[hyperindex=true,
          pdfstartview=FitH,
          bookmarksnumbered=true,
          bookmarksopen=true,
          citecolor=blue,
          linkcolor=blue,
          colorlinks=true,
          unicode]{hyperref}

\begin{document}


\title{Landau meets Newton: time translation 
symmetry breaking in classical mechanics}
\author{Liu Zhao, 
Wei Xu and Pengfei Yu\\
School of Physics, Nankai university, Tianjin 300071, China\\
{\em email}: 
\href{mailto:lzhao@nankai.edu.cn}{lzhao@nankai.edu.cn},
\href{mailto:xuweifuture@mail.nankai.edu.cn}{xuweifuture@mail.nankai.edu.cn}\\
and \href{mailto:yupengfei@foxmail.com}{yupengfei@foxmail.com}}
\date{\today}                             
\maketitle

\begin{abstract}
Every classical Newtonian mechanical system can be equipped with a nonstandard
Hamiltonian structure, in which the Hamiltonian is the square of the canonical
Hamiltonian up to a constant shift, and the Poisson bracket is nonlinear. In such a 
formalism, time translation symmetry can be spontaneously broken, 
provided the potential function becomes negative. A nice analogy between time 
translation symmetry breaking and the Landau theory of second order phase transitions is 
established, together with several example cases illustrating time translation breaking 
ground states.
In particular, the $\Lambda$CDM model of FRW cosmology is reformulated as the 
time translation symmetry breaking ground states. 
\vspace{3mm}

\vspace{3mm}

\noindent {\bf PACS: }
\end{abstract}


\section{Introduction} \label{1}

Spontaneous symmetry breaking plays an essential role in many areas in modern theoretical physics 
such as the gauge theory of particle physics and the Landau-Ginzberg 
theory of phase transitions in condensed matter physics. Spontaneously broken 
symmetries can either be internal symmetry (such as gauge and chiral 
symmetries) or spacial symmetry (such as space translation and rotation symmetries).  
Shapere and Wilczek recently discovered \cite{Shapere:2012p2114} 
that even the time translation symmetry can be 
spontaneously broken in some singular Lagrangian systems. In a 
previous work \cite{Zhao:2012p2116}, we showed that the spontaneous 
breaking of time translation symmetry 
can also be described in Hamiltonian formalism using some non-Darboux coordinates
on the phase space. We also revealed that the breaking of 
time translation is always accompanied by the breaking of time reversal.

In this work, we continue the analysis on the spontaneous breaking of time translation symmetry. We show that the phenomenon of 
spontaneous breaking of time translation symmetry can be present in almost 
every Newtonian mechanical system with a conservative potential, provided the potential 
function is shifted enough so that its value can go negative. 
The symmetry breaking occurs only in a nonstandard 
Hamiltonian description in which the Hamiltonian is the square of the canonical 
Hamiltonian and the Poisson bracket is nonlinear. Under proper 
conditions, time translation symmetry breaking can happen dynamically. 
This means that for some given set of 
initial values, time translation can be initially unbroken, but in the 
process of time,  the system can evolve into symmetry breaking phase. 
The mechanism of time translation symmetry breaking is quite 
reminiscent to the classic Landau theory of second order phase transitions,  
we just need to replace the free energy in Landau theory by the nonstandard, squared 
Hamiltonian, meanwhile take the signature of the potential 
function as order parameter (and time as an analogue of temperature). We shall make 
this analogy more concretely in the main context.
As an example, we also reformulate the $\Lambda$CDM model of cosmology as the 
time symmetry breaking ground states of a mechanical system with an 
upside-down harmonic potential.

\section{Landau theory of second order phase transition -- a mini review}
\label{2}

Landau theory of second order phase transition consists in the analysis of stable minima of the free energy as a function of some order parameter (e.g. total magnetization
$\mathcal{M}$) and the temperature $T$. In the simplest version, we take the free 
energy $F(T,\mathcal{M})$ to be of the following form,
\[
F(T,\mathcal{M})=F_{0}(T)+b(T)\mathcal{M}^{2}+ a\mathcal{M}^{4},
\]
where $a>0$ is a constant, $b(T)=b_{0}(T-T_{c})$ with $b_{0}>0$ being a constant, and 
$F_{0}(T)$ is the free energy at zero magnetization.

Stable minima of the free energy must satisfy the following conditions:
\begin{align*}
&\frac{\partial F(T,\mathcal{M})}{\partial\mathcal{M}}=0,\qquad
\frac{\partial^{2} F(T,\mathcal{M})}{\partial\mathcal{M}^{2}}>0.
\end{align*}
The first condition gives
\begin{align*}
&2\mathcal{M}(b(T) + 2a \mathcal{M}^{2})=0,
\end{align*}
from which we can get 3 distinct extrama,
\[
\mathcal{M}=0, \pm\sqrt{-\frac{b(T)}{2a}}.
\]
This, combined with the second condition, gives rise to the stable minima of the free 
energy at
\begin{align}
\mathcal{M}=\left\{
\begin{array}{ll}
0, & (T\ge T_{c}),\\
\pm\sqrt{\frac{b_{0}(T_{c}-T)}{2a}},&(T<T_{c}).
\end{array}
\right. \label{Landau}
\end{align}
At $T\ge T_{c}$, the total magnetization remains zero, indicating that the system is 
isotropic and hence rotationally symmetric. However, as $T$ drops continuously and 
reaches the regime $T<T_{c}$, the total magnetization suddenly becomes nonzero.  
This spontaneous magnetization signifies the breaking of rotational symmetry and 
hence the phase transition from the paramagnetism phase to the ferromagnetism phase.

\section{Two different Hamiltonian descriptions of Newtonian mechanics}
\label{3}

Most physicists begin their career by getting acquainted with the Newtonian equation
of motion for classical mechanics. For a unit mass particle moving in a one dimensional 
conservative potential $U(x)$, this reads
\begin{equation}
\ddot x = -\frac{dU}{dx}. \label{newt}
\end{equation}
In the standard canonical Hamiltonian formulation of classical mechanics, this equation
is obtained as the Hamiltonian flow equation arising from the canonical Hamiltonian
function
\begin{align}
H_{1}&=\frac{1}{2}v^{2}+U(x)
\end{align}
defined on the two dimensional phase space spanned by $(x,v)$ equipped with the
canonical Poisson bracket
\begin{align}
\{x,v\}_{1}&=1.
\end{align}
The flow equation actually comprises of two equations, i.e.
\begin{align}
\dot x=v,\qquad
\dot v=-\frac{dU}{dx}. \label{eqm}
\end{align}
All these are quite familiar material from every text book of classical mechanics.
What may not be so familiar comes as follows. The very same Newtonian
equation of motion (\ref{newt}), or more precisely, the canonical Hamiltonian equations 
of motion (\ref{eqm}), can also arise as the Hamiltonian equations of motion
with a completely different Hamiltonian structure. The unfamiliar Hamiltonian and 
Poisson bracket are given by
\begin{align}
&H_{2}=\left(\frac{1}{2}v^{2}+U(x)\right)^{2}+E_{0}, \label{eq1}\\
&\{x,v\}_{2}=\frac{1}{v^{2}+2U(x)},\label{eq2}
\end{align}
where $E_{0}$ is a constant, which may be chosen arbitrarily. 
This system is actually the special case $f=1/12, 
g=U$ of the $fgh$ model discussed by Shapere and Wilczek in 
\cite{Shapere:2012p2114}, but reformulated in the
Hamiltonian description given in \cite{Zhao:2012p2116}. 
The corresponding Lagrangian is
\[
\mathcal{L}=\frac{1}{12}{\dot x}^4+U{\dot x}^2-U^2.
\]
In the following, we shall be sticking exclusively to the Hamiltonian description.

Now, the same set of equations of motion (\ref{eqm}) possesses two 
different Hamiltonian descriptions, $(H_{1}, \{\cdot \,,\cdot \}_{1})$ and $(H_{2},
\{\cdot \,, \cdot \}_{2})$. Since $H_{2}=(H_{1})^{2}+E_{0}$, we see that $H_{1}$ 
and $H_{2}$ are in involution under each choices of Poisson structure. It is easy to 
check that an arbitrary linear combination of the above two Poisson structures with 
constant coefficients still makes a Poisson structure. This seems to indicate that 
every Newtonian mechanical system with a conservative force is a bi-Hamiltonian 
system. However things are not that simple. The two different Hamiltonian descriptions 
can have very different implications, as will be elucidated in the next 
section.

\section{Landau meets Newton, or ``phase transition'' in mechanical
systems} \label{4}

Consider the conditions for the Hamiltonians to have a local minimum (i.e. 
a local lowest energy state, or local ground state, which we use interchangeably below).
By the word local, we mean that such minimum may not be the absolute minimum
of the energy surface on the whole phase space. In particular, the energy surface 
may not be bounded from below even though the local minimum exists.
 
For the first Hamiltonian $H_{1}$, the conditions for the Hamiltonian to have a local minimum read
\begin{align}
&v^{2}=0,\quad\frac{dU}{dx}=0, \quad \frac{d^{2}U}{dx^{2}}>0,
\end{align}
i.e. at the local ground state, the particle must be at rest and the potential must take 
its minimum. If $U(x)$ doesn't have a minimum, then the Hamiltonian does not have a 
local ground state, and is unbounded from below. Whenever the Hamiltonian does 
have a local ground state, the state is kept by time translation 
symmetry (since both $x$ and $v$ are fixed in time
in such a such state). Note that any such state must be isolated, 
though it may not be unique (except the most trivial case with a constant potential).

Now consider the second Hamiltonian description given by eqs.(\ref{eq1}) and 
(\ref{eq2}). The conditions for the energy to have a local minimum are
\begin{align}
v^{2}&=0,\quad \frac{dU}{dx}=0, \quad  \frac{d^{2}U}{dx^{2}}>0, 
&& \text{if $U(x)\ge 0$,} \label{grd1}\\
v^{2}&=-2U(x) && \text{if $U(x)< 0$. } \label{grd2}
\end{align}
In the former case, the local minimum of the energy is in general higher than 
$E_{0}$ (since $U(x)$ needs not to be zero at its minimum), and 
time translation is unbroken. Such local ground states are similar in spirits to
the case which we encountered in the first Hamiltonian description. In the latter case, 
the minimum energy is equal to $E_{0}$, the force $-\frac{dU}{dx}$ is not 
necessarily zero, and 
time translation is broken because of the nontrivial motion in the ground state.  This 
kind of ground states are not isolated (they actually form two smooth curves on the 
phase space), and they never appear in the first Hamiltonian description. 
Therefore, we observe significant differences between the two descriptions: 
\begin{itemize}
\item Though we use the term ``potential function'' for $U(x)$ in both Hamiltonian 
descriptions, the actual roles of $U(x)$ in the two descriptions are quite different. In 
particular, $U(x)$ in the first Hamiltonian has the well known interpretation as potential 
energy, while $U(x)$ in the second description does not have such a simple 
interpretation. It does not even have the same dimensionality with the energy;
\item A constant shift in the potential is physically insignificant in the first description, 
while it becomes significant in the second description;
\item Given the same potential function $U(x)$, the energy is not necessarily bounded 
from below in the first description, while it is always bounded from below in the second 
description. Though for classical mechanical systems, the existence of a lower energy 
bound looks insignificant, it immediately becomes significant when passing to 
quantum description is under consideration;
\item When the energy possesses local minima, time translation is always
preserved in the local ground states in the first description, while it can be broken 
in the second description as long as the potential becomes negative.
\end{itemize}

One can compare the breaking of time translation in the second description 
to the case of Landau theory of second order phase 
transitions reviewed previously. Both theories involve 
symmetry breaking. In Landau theory, the order parameter is the spontaneous 
magnetization $\mathcal{M}$, while in our case, the velocity $v$
plays a similar role. Moreover, in Landau theory, the order parameter is intrinsically 
controlled by the temperature, while in the present case, the value of $v$ at the ground
state depends intrinsically on the value of $U(x)$, and $U(x)$ is controlled by $x$ and 
hence indirectly by $t$.
 
It is necessary to have some intuitive feelings about the time translation symmetry 
breaking mentioned above. For this purpose let us consider some simple 
example cases.

The first example case is the constant potential case $U(x)=-\frac{1}{2}u^{2},\, (u>0)$. 
In the symmetry breaking lowest energy state, we have $v=\pm u$. Of course, the 
particle can pick only one of the two values for the velocity at one time. The explicit 
choice would then break the time reversal.

The second example case is given by a linear potential, $U(x)=- F x$. If the particle
was initially sitting at $x=x_{0}\leq 0$, it will be driven by the constant force $F$ 
towards the positive direction along the $x$ axis until it crosses the origin. 
In particular, if the particle were released from at rest at the origin, then its motion 
will be confined in the lowest energy states governed by the equation $v^{2}=2Fx$. 

In the third example case we consider the quadratic potential. We can actually have
two different choices of quadratic potentials: the usual harmonic oscillator potential with 
a constant shift, $U(x)=\frac{1}{2}k x^{2}-U_{0}$, and the upside-down harmonic
potential,  $U(x)=U_{0}-\frac{1}{2}k x^{2}$. 
For both choices the symmetry breaking lowest 
energy states are well defined and they all corresponds to nontrivial motion. 
The upside-down harmonic oscillator has appeared in the studies of 
matrix description of de Sitter gravity \cite{Gao:2001p2174} and the Hamiltonian 
formalism of FRW cosmology \cite{Ren:2007p4461}. We will 
come back to this model in Section \ref{5}.

Of course almost everyone is familiar with the above potential within the framework of 
the first Hamiltonian description. There, the second example and the model with 
the upside-down harmonic potential in the third example do not have a lower 
energy bound. The first example has a flat potential, and though the lowest energy 
states do exist, they do not subject to nontrivial motion. The model with the harmonic 
potential in the third example also possesses physically well behaved lowest energy state, 
which also does not subject to nontrivial motion.

Now back to the second Hamiltonian description. We can choose arbitrary, more 
complex potentials and see whether the time translation symmetry breaking 
can occur in a dynamical process. Let us consider, for instance, the potential
function 
\[
U(x)=-x^3 + 5 x^2 + 7,
\]
which is chosen at random. Figure 1 gives the plot of the energy surface 
in the two dimensional phase space spanned by $(x,v)$ 
(where we have chosen $E_{0}=0$). The energy surface has an 
isolated local minimum (whose value is above zero) at $(x,v)=(0, 0)$, 
which is the time translation preserving ground state of the system, as well as a whole 
curve of absolute minima (whose value is exactly zero) to the right of the plot, which 
correspond to the time translation breaking ground states. 

\begin{figure}[ht]
\begin{center}
\includegraphics[width=0.7\textwidth]{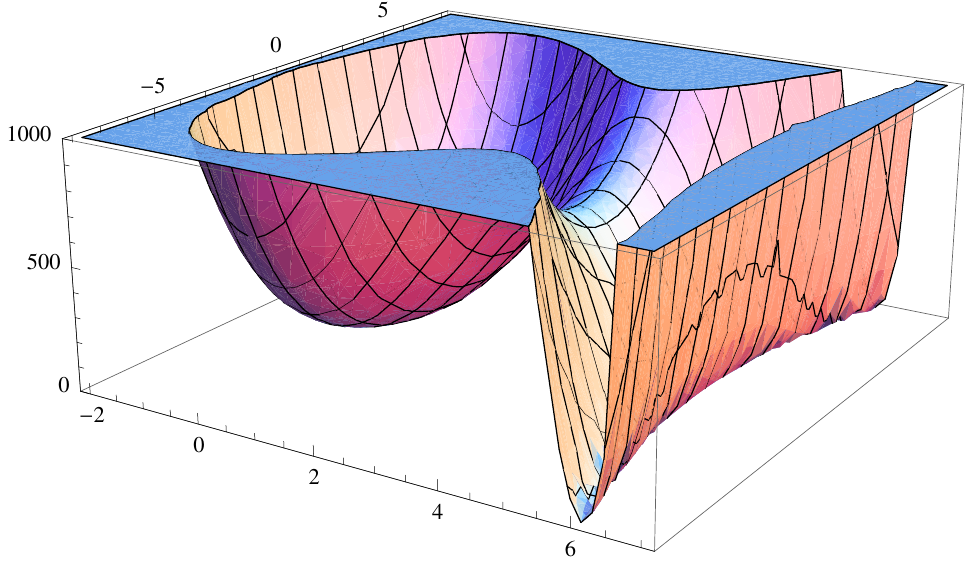}
\begin{minipage}{0.9\textwidth}
\caption{An illustration of the energy surface with the potential function 
$U(x)=-x^3 + 5 x^2 + 7$ in the second Hamiltonian description}
\end{minipage}
\end{center}
\end{figure}

Now if the particle were initially sitting at the isolated, symmetry preserving, local 
ground state, can it evolve into the symmetry breaking ground states in the course of 
time? This looks impossible because 
the particle does not have enough energy 
to climb up the energy barrier between the isolated local ground state and the curve of 
the absolute minima of the energy surface. However, there are at least two 
possible mechanisms which can make the particle to transit from the symmetry 
preserving ground state to the symmetry breaking ground states. The first possibility 
is to allow the particle to be not alone in the system. It can be part of a larger system
and hence can be perturbed by other degrees of freedom in the larger system so that it 
gets enough energy to climb the energy barrier and then loses energy by other 
perturbations and stay in the symmetry breaking ground states henceforth. Another 
possibility is to consider quantum effect. If the particle's motion is controlled by 
quantum principles, then it has some possibility of penetrating the energy barrier 
by quantum tunneling process. In either way the breaking of time translation symmetry 
can happen in a dynamical process. It is of particular interests if the particle can carry 
some kind of charge, because then the transition from the isolated local ground state to 
the continuous symmetry breaking ground states would correspond to a  
transition from insulating phase to a superconducting phase for the appropriate charge. 
In this sense, the analogy to the Landau theory of second order phase transitions may 
not seem to be an accident.

\section{An application: $\Lambda$CDM model of the universe as symmetry breaking ground states}
\label{5}

Perhaps the most important physical system in which time translation symmetry is 
explicitly broken is the Friedmann equations governing the evolution of our universe. 
In modern treatments, the Friedmann equations arise as components of 
the Einstein equation of general relativity with the insertion of the FRW metric. 
However, in some approximate, or effective description, they are also known to arise
from Newtonian mechanics \cite{Milne}, as reviewed, e.g. in \cite{Longair}.
Below we shall try to incorporate Friedmann equations into the framework 
of the second Hamiltonian description of Newtonian mechanics. In particular, 
the first order equation in the Friedmann equations will appear as the condition 
determining the time translation symmetry breaking ground states.

Recall that the Friedmann equations actually consist of two equations, i.e.
\begin{align}
&\left(\frac{\dot a}{a}\right)^{2}+\frac{k}{a^{2}}-\frac{\Lambda}{3}
=\frac{8\pi G\rho}{3}, \label{f1}\\
&\frac{\ddot a}{a}-\frac{\Lambda}{3}=-\frac{4\pi G}{3}(\rho+3p), \label{f2}
\end{align}
where $a(t)$ is the FRW scale factor, $G$, $k$, $\Lambda$, $\rho$, $p$  are 
respectively the Newtonian constant for gravitation, spacial curvature, cosmological 
constant, density and pressure of the ideal fluid source of the universe. The presence 
of $\rho$ and $p$ makes it difficult to think of the Friedmann equations as the equations 
of motion of a purely mechanical system. So, we will consider only the 
sourceless Friedmann equations, i.e. the $\Lambda$CDM model.

The incorporation can be made quite easily. We just need to rearrange the equations (\ref{f1}) and (\ref{f2}) with $\rho=p=0$ into the following form,
\begin{align}
&\dot a^{2}=-k+\frac{\Lambda a^{2}}{3}, \label{fr1}\\
&\ddot a = \frac{\Lambda a}{3}. \label{fr2}
\end{align}
Then, by identifying $a(t)=x(t), v(t)=\dot a(t)$ and making comparison between 
(\ref{fr1}) and the symmetry breaking ground state condition (\ref{grd2}), we 
immediately realize that the above system is the very same model with upside-down 
harmonic potential in the third example of 
Section \ref{4} provided $\Lambda>0$, which is required for a 
de Sitter universe. Explicitly, we now have
\begin{align*}
&U(a) = \frac{k}{2}-\frac{\Lambda a^{2}}{6},
\\
&H_{2}=\left(\frac{\dot a^{2}}{2}
+ \frac{k}{2}-\frac{\Lambda a^{2}}{6}\right)^{2},
\end{align*}
and (\ref{fr2}) is just the Newtonian equation of motion that follows from this 
model.

Let us proceed to see what we can learn by reinterpreting the equations (\ref{fr1}) and
(\ref{fr2}) respectively as the symmetry breaking ground state condition and the 
Hamiltonian equation of motion associated with our second Hamiltonian description of
Newtonian mechanics. 

The first lesson we learn from above is that although the universe could undergo
perpetual time evolution (since $\dot a \neq 0$), it remains in the lowest energy state
$H_{2}=0$. The second lesson is that the spacial curvature, $k$, has to be 
non-positive (and preferably be zero) if the 
universe began its evolution from a very small size which is close to zero. 
Otherwise the ground state condition,
(\ref{fr1}), will not have a solution, and our formulation of the $\Lambda$CDM
model as the symmetry breaking ground state collapses. 
The use of the second Hamiltonian description feels more natural
than using the first Hamiltonian description (as was previous did in 
\cite{Ren:2007p4461}), because 
only the zero energy condition seems to be consistent with the general 
relativistic Hamiltonian constraints and with the formalism of 
Wheeler-deWitt equation (though with a different Hamiltonian).

In closing this section, let us remark that the formalism we introduced here for
describing $\Lambda$CDM model is only an effective description. The fundamental 
description should still be based on relativistic theory of gravity, preferably 
on some higher curvature generalizations of general relativity. It will be very interesting 
if our second Hamiltonian could arise from the Legendre transform of the reduced 
Lagrangian of some generalized gravity model after the insertion of FRW metric
into the action.

\section{Conclusions}

Time translation symmetry breaking seems to happen only in systems with 
a higher order Lagrangian or Hamiltonian. In this work, we are particularly interested in 
a special kind of Hamiltonians which yield spontaneous breaking of time translation.
The time evolution equations for such systems are identical to the Newtonian equation
with a conservative potential. The Hamiltonian is the square (up to a constant shift) of 
the standard canonical Hamiltonian for Newtonian mechanical systems, which is 
equipped with a nonlinear Poisson structure. 

For such systems, the spontaneous breaking of time translation is quite reminiscent to
the mechanism of symmetry breaking appearing in Landau theory of second order phase 
transitions. We gave the detailed analogy in the main context. We also sketched some
possible mechanisms which can make the breaking of time translation happen 
dynamically. Besides all these, we gave 
several example cases indicating the explicit breaking of time translation. In particular, 
the $\Lambda$CDM model of FRW cosmology is reformulated as time translation
breaking ground states for a system with upside-down harmonic potential. 

The works reported in \cite{Zhao:2012p2116} and in the present paper are still
very preliminary. There are still a lot of open issues which need further studies. 
Among these, we would like to point out a few problems which we would like to
study in subsequent works:
\begin{itemize}
\item Within the context of Hamiltonian description, the passage to the quantum 
description is a much needed whilst still missing piece of work. Although path integral
quantization may be a possible choice \cite{Anonymous:2012p1901}, 
the analogue of canonical quantization is still needed 
in order to have a quantum mechanical description for the models under consideration;
\item The unusual relationship between the potential function $U(x)$ and the 
Hamiltonian $H_{2}$ in the second Hamiltonian description needs further
understanding. It indicates that the usual, canonical way of introducing interactions
is not the only possible choice;
\item  We have been exclusively considering the problem of time translation breaking
within the scope of classical mechanics. It will of course also be interesting if similar 
mechanisms can also arise in field theoretic context.
\end{itemize}

\section*{Acknowledgment} 

This work is supported by the National  Natural Science Foundation of 
China (NSFC) through grant No.10875059. L.Z. would like to thank 
X.-H. Meng, H.-X. Yang and the 
organizer and participants of ``The advanced workshop on Dark Energy 
and Fundamental Theory'' supported by the Special Fund for Theoretical 
Physics from the National Natural Science Foundation of China with grant 
no: 10947203 for useful discussions.


\providecommand{\href}[2]{#2}\begingroup\raggedright\endgroup

\end{document}